# A Low Power and Precise Tunable Voltage Amplifier Based on the Translinear Circuit Scheme of CCCII+

**Umar Mohammad**

Assistant Professor, Department of Electronics & Communication Engineering
Islamic University of Science & Technology, Awantipora, Kashmir, 192122
e-mail: umarnaik@iul.ac.in

In the field of analog circuit design, current conveyors have equitably established their uniqueness as an important circuit design element. The literature available to us during the few years in the field of analog VLSI design, quotes a huge number of application elements based on current conveyors. Likely, in this paper, a new tunable voltage amplifier based on the translinear circuit scheme of second generation current controlled current conveyor has been proposed. The modeling of the circuit presented in this paper employs the minimum number of passive elements. The magnitude of the tuning or the amplitude of the voltage presented here, is being controlled by means of two variable resistors. Current conveyor second generation translinear circuit scheme is taken into consideration to implement the proposed tunable voltage amplifier. CCCII works on the outlines of low power and low voltage design. Tunable voltage amplifiers find use in analog as well as in digital signal processing applications.

## 1 INTRODUCTION

Current conveyors [1] during the past few years have revolutionized the analog circuit design application area. Due to the low power and low voltage characteristics, the translinear circuit schemes of the current conveyors have proven worthy enough too. Secondly, the Current mode (CM) circuit design approach is firm attaining and establishing a trend setting reputation in the field of modern day VLSI. It proves to be a unique technique, which can help applying various design considerations which are ineffective or almost impossible to apply otherwise. In other terms its superiority over the voltage mode approach is trend setting [2].

Current Conveyor is a three terminal device, in which the input/output terminal are able to convey both current and voltage to each other. CCI, CCII, CCIII are the three generation of the current conveyors proposed by the Sedra, smith (CCI, CCII) [3] and Fabre (CCIII) [4]. The use of CC's doesn't increase the complexity of the overall circuit designed. In this work, we are using the second generation current controlled current conveyor, because of the outstanding features like availability of band width, auto balancing, optimization of hardware, no triggering required etc.

A simple Voltage amplification scheme using CCCII+ is presented in this paper. The advantage of this type of amplifier is that, it requires a single CCCII+ circuit and two resistors, to amplify the input signal. The work presented in this paper has been Simulated and verified using HSPICE tool. The 45nm predictive modelling Parameters of Mosfet have been taken from the Arizona State University [5].

## 2 CCCII+ AND DESIGN CONSIDERATIONS

The scheme, we have adopted in the second generation current conveyor is the current mirror based translinear circuit (CCCII+) [6] shown below in figure 2. The advantage of using this circuit scheme is that it is fully balanced. The intrinsic resistance in this type can be balanced by tuning the biasing current $I_b$.

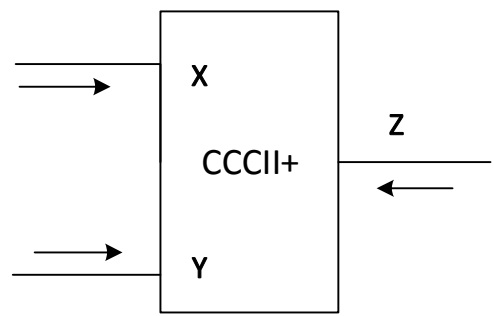

**Figure. 1:** General Representation of CCCII+

The matrix representation of CCCII is below;

$$\begin{bmatrix} V_X \\ I_Y \\ I_Z \end{bmatrix} = \begin{bmatrix} R_X & 1 & 0 \\ 0 & 0 & 0 \\ \pm 1 & 0 & 0 \end{bmatrix} \begin{bmatrix} I_X \\ V_Y \\ V_Z \end{bmatrix}$$

**Figure. 2:** Matrix Representation of CCCII± **[13]**

From the above matrix, we can conclude the following things relating the operation of CCCII±.

$$I_Y = 0$$
$$V_X = V_Y + I_X . R_X$$
$$I_Z^+ = + I_X$$
$$I_Z^- = -I_X$$

Resistance at $X = R_X = 1/2g_m = 1/\sqrt{8.\beta_n.I_B}$,

Where $\beta_n = \mu_n C_{ox}.W/L$

The above equations are well defined from the above matrix in figure 2, in which we have, **"$I_Y$"**--node current at Port **"Y"**, **"$V_X$"** --voltage at node "X", $V_Y$ --voltage at node **"Y"**. **"$I_Z$"** and **"$V_Z$"** are the current and Voltages at the node **"Z"** respectively. **"$R_X$"** is the intrinsic resistance of the translinear circuit.

Some of the promising features of the CCCII are **[7], [8]**;

1. Greater Linearity
2. Availability of full bandwidth
3. Autobalaing at Higher temperatures
4. Low Power, Low Voltage

### 3 FERRI, GIUSEPPE, AND NICOLA C. GUERRINI VOLTAGE AMPLIFIER USING CCII.

In literature, we have CCII, based voltage amplifier, proposed by Ferri, Giuseppe, and Nicola C. Guerrini **[9]**. In this circuit, there is a requirement of two CC's and two passive elements. The" Z" node in one of the CC's has been put in floating mode as shown in figure 3 below.

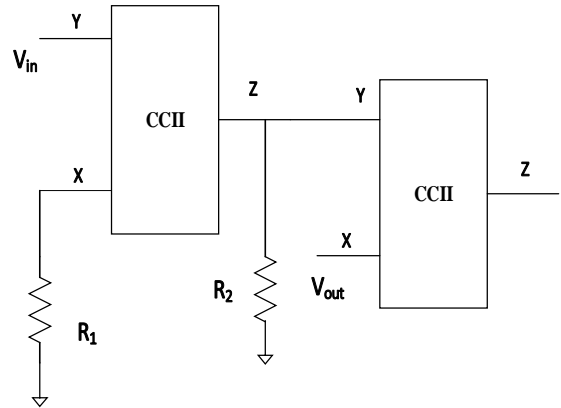

**Figure 3.** Ferri, Guerrini, CCII based Voltage Amplifier **[9]**.

### 4 PROPOSED TUNABLE VOLTAGE AMPLIFIER

A variety of applications based on CCCII are available **[10] [11]**. But, as per the author's prior knowledge, voltage amplifier based on the scheme of CCCII+ is not available yet. However, a voltage amplifier employing one and two CCII's in available **[12]**. The proposed circuit using CCCII+ performing Voltage amplification is provided below in figure 4, along with its voltage gain relationship. The advantage of using this type of current conveyor is the requirement of less hardware, low power consumption and less complexity.

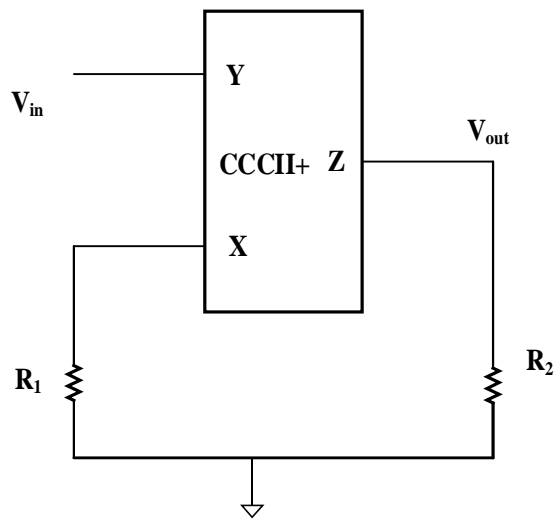

**Figure 4.** Proposed Tunable Voltage amplifier

The voltage gain relationship for the circuit proposed in figure 4, can be concluded as;

$$V_{out} = R_2.I_Z = R_2.I_X = R_2.\frac{V_X}{R_1}$$
$$V_{in} = V_Y = V_X$$
$$V_{out} = \frac{R_2}{R_1} V_{in}$$

The nomenclature of the circuit provided in figure 4 can be summarized as; At X, Y and Z are the three terminals of the CCCII+. $V_{in}$ is the input voltage at node X, $V_{out}$ is the output of the CCCII+ at node Z **[13][14]**. $R_1$ and $R_2$ are the variable resistances provided at the nodes Y and X. The change in these two parasitic resistances are responsible for the increase and decrease in the amplitude of the input signal provided at node X. An increase in the value of the resistances provided, a gradual decrease in the amplitude of the Input signal is seen. Similarly lowering the values of the resistances, we can observe a sharp increase in the amplitude of the input signal. The translinear circuit scheme used in this Current conveyor is shown below;

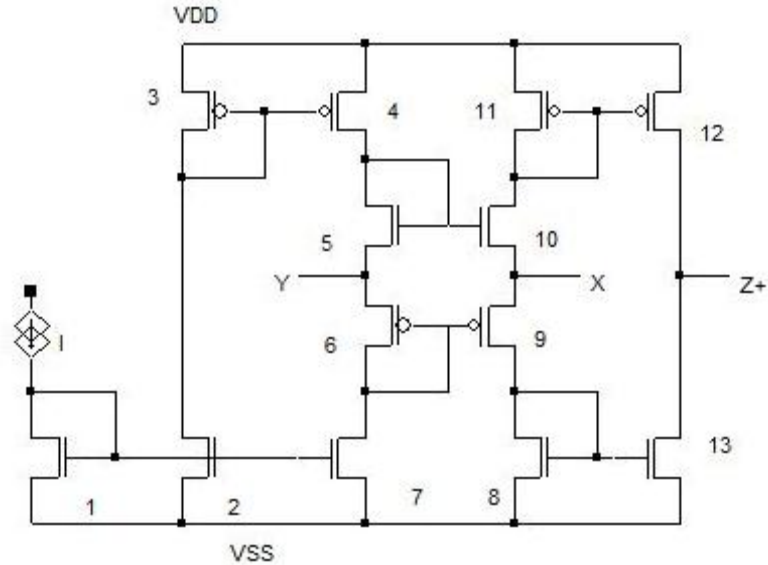

**Figure 5. Translinear circuit used for CCCII+**

## 5 Simulation Results:

The device analysis results obtained from the simulation of the comparative and the proposed circuit schemes provided in this work is viz;

| Circuit | No. of CC's | No. of Resistors | Technology | Floating nodes | Average Power Dissipation | Peak Power |
|---|---|---|---|---|---|---|
| Ferri, Guerrini Voltage Amplifier[12] implemented using CCII+ | 2 | 02 | 45nm PTM CMOS | 01 | 1.3154E-05 | 1.3178E-05 |
| Ferri, Guerrini Voltage Amplifier[12] implemented using CCII+ | 01 | 02 | 45nm PTM CMOS | nil | 4.9500E-06 | 4.9561E-06 |
| Ferri, Guerrini Voltage Amplifier implemented using CCCII+ | 02 | 02 | 45nm PTM CMOS | 01 | 2.1233E-04 | 2.2915E-04 |
| **Proposed Voltage Amplifier using CCCII+** | 01 | 02 | 45nm PTM CMOS | nil | 1.1905E-04 | 1.2388E-04 |

**Table. I. Results of the Tunable Voltage Amplifier.**

Different aspects of amplification are seen on the simulating tool, while tuning the resistors $R_1$ and $R_2$. Such behaviors have been discussed in the Table 2. The tuning of resistances should be done as per the proper guideline shown under Case II in the table to for the proper amplification.

| S.no | $R_1$ Vs $R_2$ | Output Behaviour |
|---|---|---|
| **Case I** | $R_1$ greater than $R_2$ | Input signal amplitude is decreased |
| **Case II** | $R_2$ greater than $R_1$ | Proposed Circuit Behaves as Tunable Voltage Amplifier. |
| **Case III** | $R_1$ equals $R_2$ | Input signal amplitude is decreased |

**Table II. Scheme of the Tunable resistors in the Proposed Voltage Amplifier.**

**Simulated results of amplification on HSPICE Tool:**

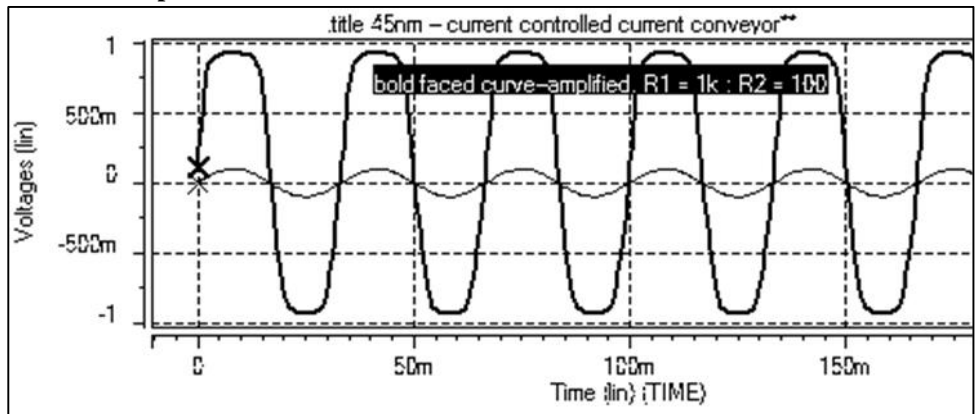

**Figure 6. Voltage Amplification, R1 = 1k and R2 = 100k, $V_{in}$ = 100mV$_{PP}$: $V_{out}$ = 1 $V_{PP}$**

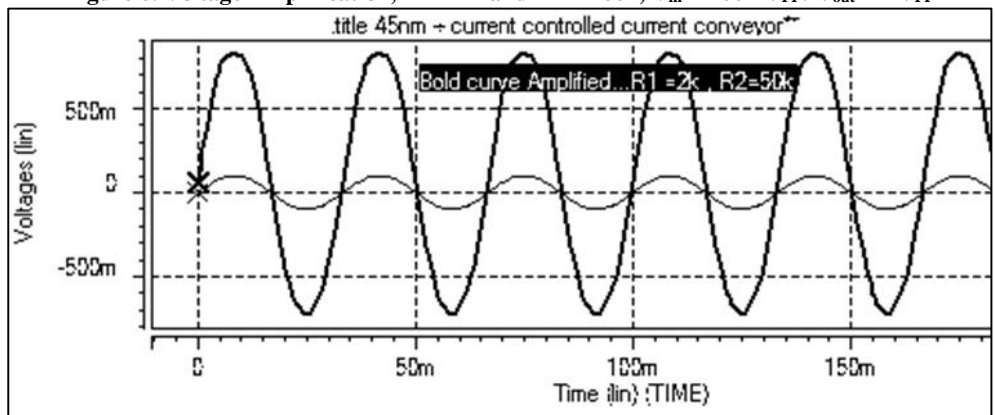

**Figure 7. Voltage Amplification, R1 = 2k and R2 = 50k, Vin = 100mV; $V_{out}$ = 800mV$_{PP}$**

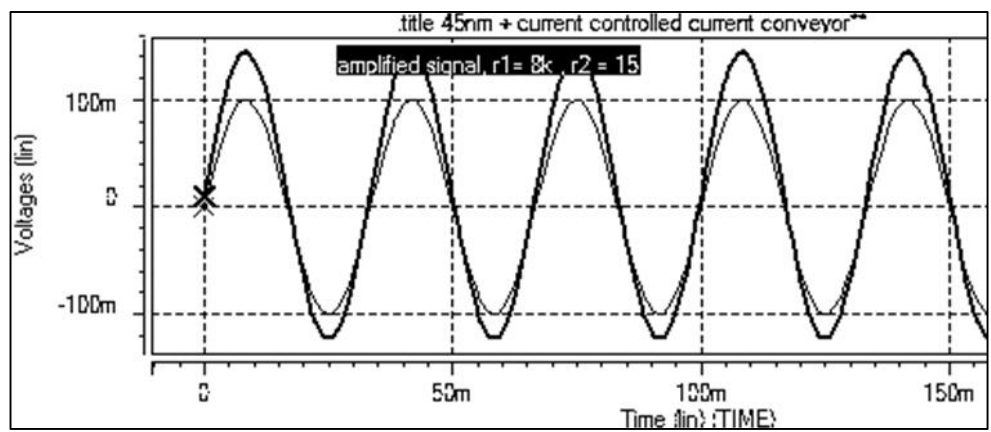

**Figure 8. Voltage Amplification, R1 = 8k and R2 = 15k, $V_{in}$ = 100mV; 130mV$_{PP}$**

# 6 CONCLUSION

A novel tunable voltage amplifier employing less hardware is presented in this paper. The proposed voltage amplifier works on the scheme of the Active device Current Conveyor, through a generalized and distinct translinear circuit. The simulation results of Proposed circuit were obtained in HSPICE tool in the 45nm Predictive model Mosfet Technology. The evaluation results of device parameters among the Power, PDP delay, Peak Power, etc. have also been discussed in the paper.

## ACKNOWLEDGEMENT

The work presented in the paper was carried out at the Signal Processing Lab, Department of Electronics, Islamic University of Science & Technology, Kashmir-192122. The authors would like to thank the staff of the Lab for their kind Support and Patience.

# REFERENCES

1 Smith, K. C., Sedra, A. S., "The Current Conveyor: A New Circuit Building Block," Proc. IEEE, pp. 1368_1369 (1968). doi: 10.1109/PROC.1968.6591

2 Yasin, Mohd Yusuf, and Bal Gopal. "High-frequency oscillator design using a single 45 nm CMOS current controlled current conveyor (CCCII+) with minimum passive components." Circuits and systems 2, no. 02 (2011): 53.

3 Smith, K., and A. Sedra. "Realization of the Chua family of new nonlinear network elements using the current conveyor." IEEE Transactions on Circuit Theory 17, no. 1 (1970): 137-139.

4 Fabre, Alain. "Third-generation current conveyor: a new helpful active element." Electronics Letters 31, no. 5 (1995): 338-339.

5 CMOS model parameters-Predictive Technology Model, 2006, http://ptm.asu.edu.

6 Sedra A., Smith K., "A second-generation current conveyor and its applications", IEEE Trans., vol. CT-17, pp 132-134, 1970.

7 Senani, Raj, D. R. Bhaskar, and A. K. Singh. Current conveyors: variants, applications and hardware implementations. Springer, 2014.

8 M.Y.Yasin, "Power minimization of analog circuits in deep submicron technology", Ph.D.THESIS, Integral University Lucknow, 2013.hdl.handle.net/10603/19775.

9 Ferri, Giuseppe, and Nicola C. Guerrini. Low-voltage low-power CMOS current conveyors. Springer Science & Business Media, 2003.Page-141.

10 Minaei, Shahram, Onur Korhan Sayin, and Hakan Kuntman. "A new CMOS electronically tunable current conveyor and its application to current-mode filters." IEEE Transactions on Circuits and Systems I: Regular Papers 53.7 (2006): 1448-1457.

11 Maheshwari S. High CMRR wide bandwidth instrumentation amplifier using current controlled conveyors. International Journal of Electronics. 2003 Dec 31;89(12):889-96.

12 Ferri, Giuseppe, and Nicola C. Guerrini. Low-voltage low-power CMOS current conveyors. Springer Science & Business Media, 2003.Page-27,141.

13 Abbas, Zia, Giuseppe Scotti, and Mauro Olivieri. "Current controlled current conveyor (CCCII) and application using 65nm CMOS technology." World Academy of Science, Engineering and Technology55 (2011): 935-939.

14 Fabre, A., Saaid, O., Wiest, F., & Boucheron, C. (1996). High frequency applications based on a new current controlled conveyor.IEEE Transactions on Circuits and Systems I: Fundamental Theory and Applications, 43(2), 82-91.